\documentclass[]{jkas}

\usepackage{graphicx}
\usepackage{hyperref}

\def\year{2014} 
\def\month{March} 
\def\volume{00} 
\def\issue{0} 
\def\beginpage{1} 
\def\endpage{99} 
\def\received{---} 
\def\accepted{---} 
\date{Received \received ; accepted \accepted}

\title{Moreton Waves Related to the Flare Event\\ in 3 June 2012 and 6 July 2012}
\author[1]{Agustinus Gunawan~Admiranto}
\author[1,2]{Rhorom~Priyatikanto}

\affil[1]{National Institute of Aeronautics and Space, Indonesia; \email{gun\_agustinus@yahoo.com}}
\affil[2]{Astronomy Program, Institut Teknologi Bandung, Indonesia; \email{rho2m@hotmail.com}}

\def\corrauthor{A. G. Admiranto}
\def\runningauthor{A. G. Admiranto \& R. Priyatikanto}
\def\runningtitle{Moreton Waves -- 2012-06-03 \& 2012-07-06}
\def\keywords{Sun: activity --- Sun: chromosphere --- Sun: flares}

\newcommand{\halp}{H$\alpha$}

\def\abstracttext{
In this study, we present geometrical and kinematical analysis of Moreton wave observed in 2012 June 3rd and July 6th, recorded in H-$\alpha$ images of Global Oscillation Network Group (GONG) archive. These large-scale waves exhibit different features compared to each other. The observed wave of June 3rd has angular span of about $70^{\circ}$ with a diffuse wave front associated to NOAA active region 11496. It was found that the speed of the wave that started propagating at 17.53 UT is about $931\pm80$ km/s. The broadness nature of this Moreton wave can be interpreted as the vertical extension of the wave over the chromosphere. On the other hand, the wave of July 6th may be associated to X1.1 class flare that occurred at 23.01 UT around the 11515 active region. From the kinematical analysis, the wave propagated with the initial velocity of about $994\pm70$ km/s which is in agreement with coronal shock velocity derived from type II radio burst observation, $v\sim1100$ km/s. These two identified waves add the inventory of the large-scale waves observed in 24th Solar Cycle.}

\begin{document}
\jkashead

\section{Introduction}
Moreton waves are a kind of large scale atmospheric wave which occur in the solar chromospher. This kind of wave was first observed through H$\alpha$ filter by \cite[Moreton][]{moreton60} and have velocities in the range of 500-1500 km/s, and the angular extent of 90-270 degrees \citep{warmuth04,bala10}. This kind of waves are often associated with flare or coronal mass ejection (CME) which are considered as the trigger of the waves \citep{uchida73,chen02,temmer09}. \cite[Uchida][]{uchida73} proposed the "sweeping skirt" hypothesis, interpreted Moreton waves as the rapidly moving intersection of the chromosphere and the flare produced coronal fast-mode wavefront. In this nature, Moreton waves are closely related to its coronal counterpart namely EIT waves \citep{thompson99} that move with signficantly lower speed (2-3 times slower), accompanied type II radio burst as the result of the rising plasma.

In this paper, we report the Moreton waves which occured on June 3, 2012 and July 6, 2012 as recorded in series of H$\alpha$ images obtained from the Global Oscillation Network Group \citep[GONG][]{hill94} data archive. Kinematical and geometrical analyses are expected to provide some insights about the physical mechanisms and relationship with other solar atmospheric phenomena, i.e. flares, coronal mass ejections, and propagation of EIT waves.

\section{Data and Method}
\begin{figure}[ht]
\centering
\includegraphics[width=0.12\textwidth]{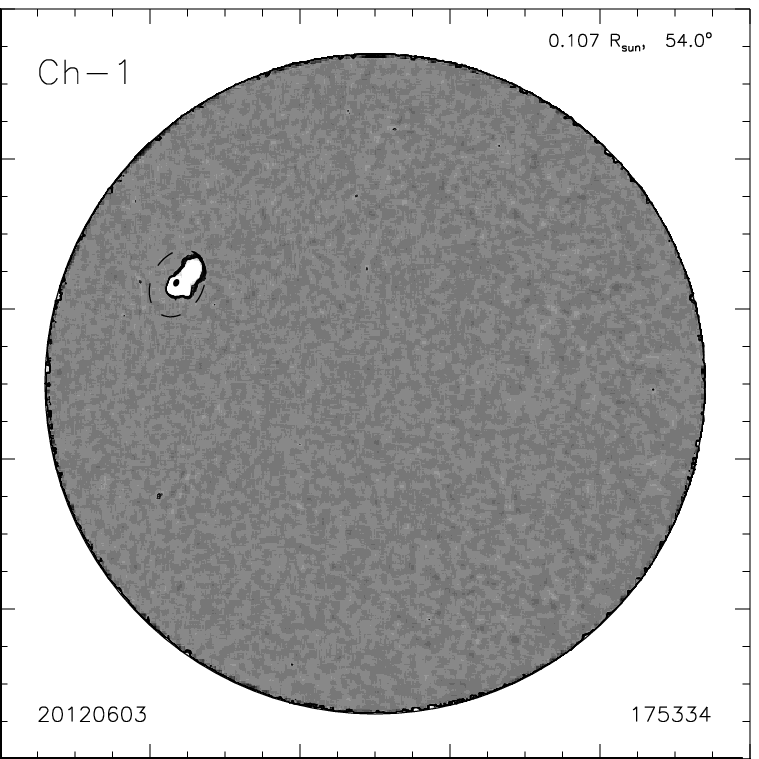}\hfill
\includegraphics[width=0.12\textwidth]{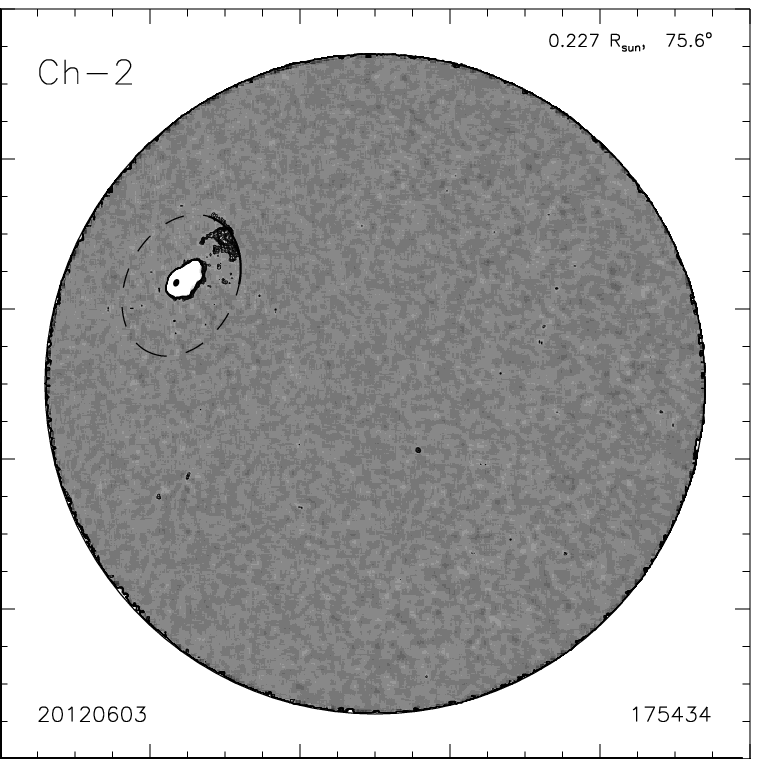}\hfill
\includegraphics[width=0.12\textwidth]{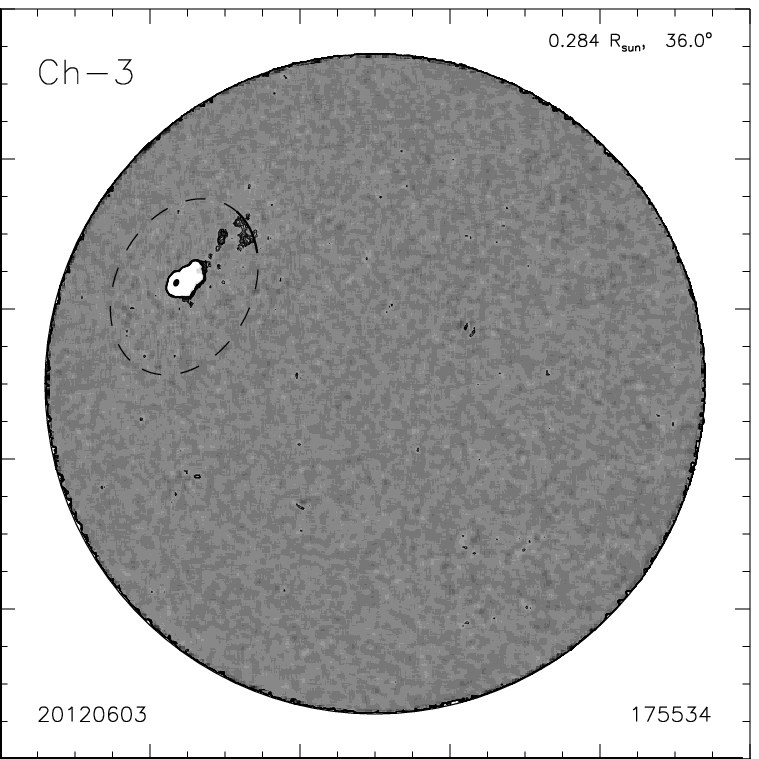}\hfill
\includegraphics[width=0.12\textwidth]{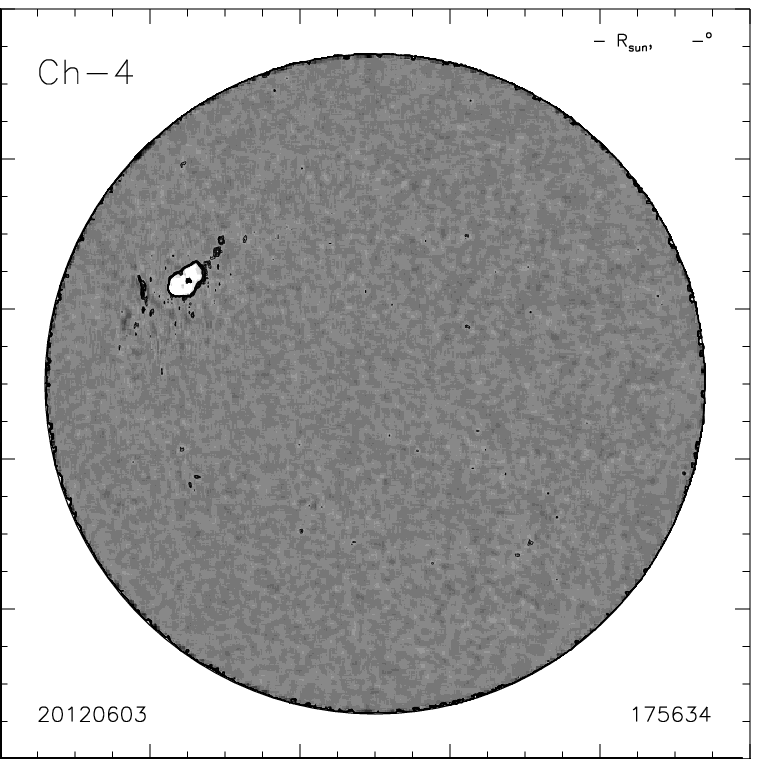}\\
\vskip5pt
\includegraphics[width=0.12\textwidth]{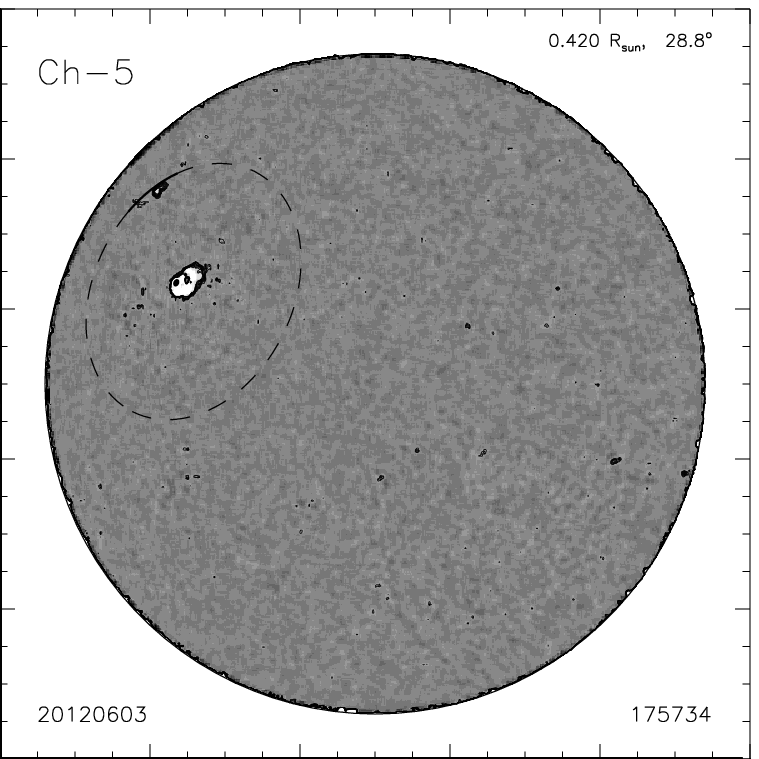}\hfill
\includegraphics[width=0.12\textwidth]{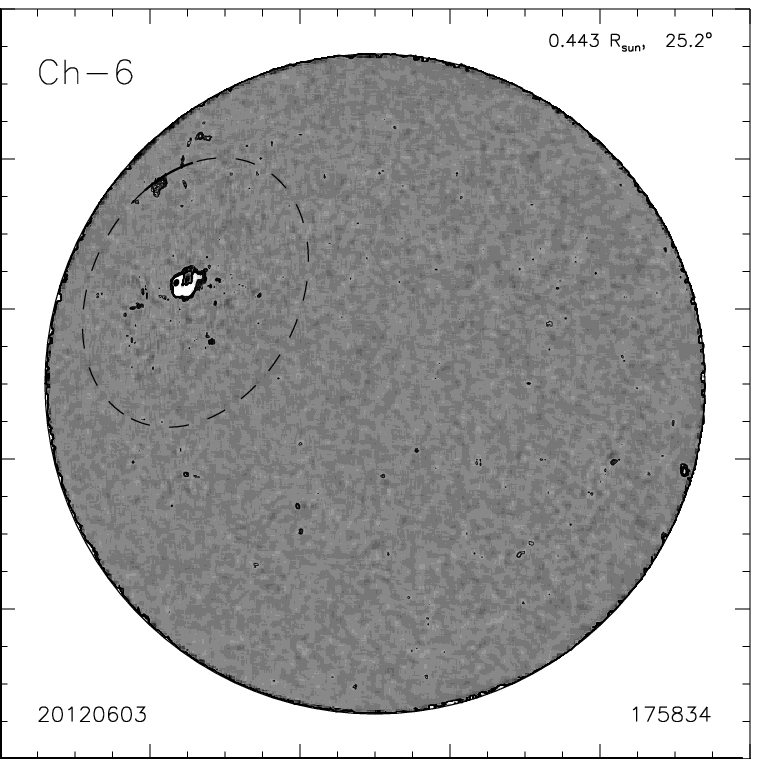}\hfill
\includegraphics[width=0.12\textwidth]{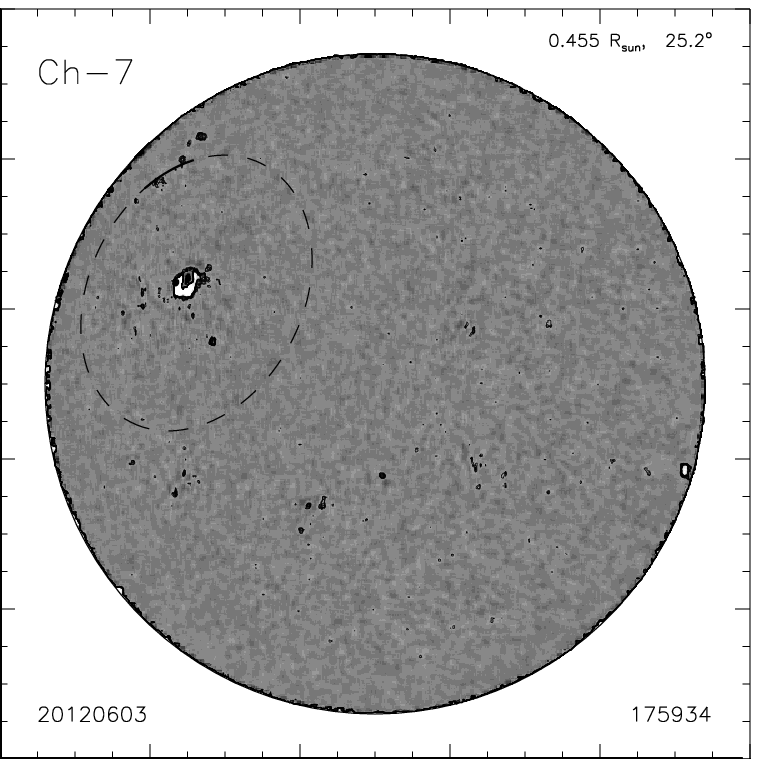}\hfill
\includegraphics[width=0.12\textwidth]{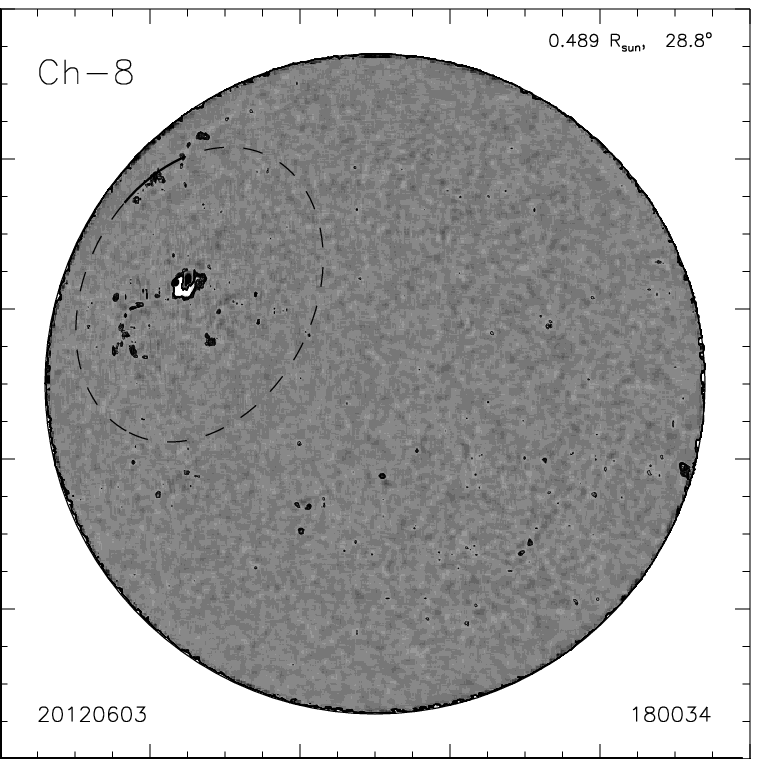}\\
\caption{Series of H$\alpha$ diffrent images (after contrast adjustment) of the June 3 event observed in Cerro Tololo Observatory as obtained from \url{nso2.gong.com}. Date and time of observation are displayed in each frame. Contour lines are superposed to the images to emphasize the shape of Moreton wavefronts. Circular arcs with certain radial distance from the wave origin are fitted manually. Lower panels show the brightening feature observed in the north-eastern part of AR 1496.}
\label{halpha}
\end{figure}

\begin{figure}[ht]
\centering
\includegraphics[width=0.12\textwidth]{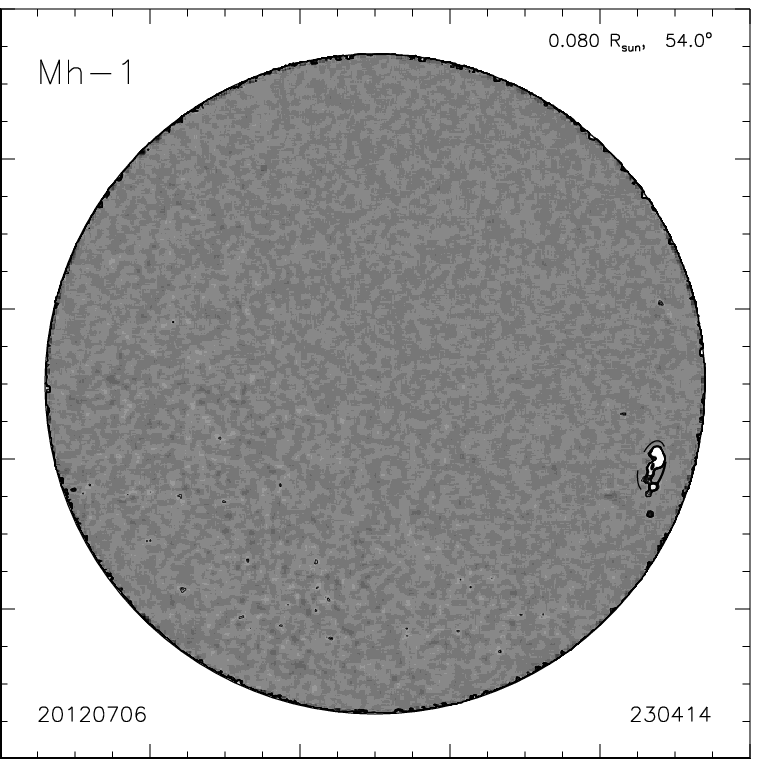}\hfill
\includegraphics[width=0.12\textwidth]{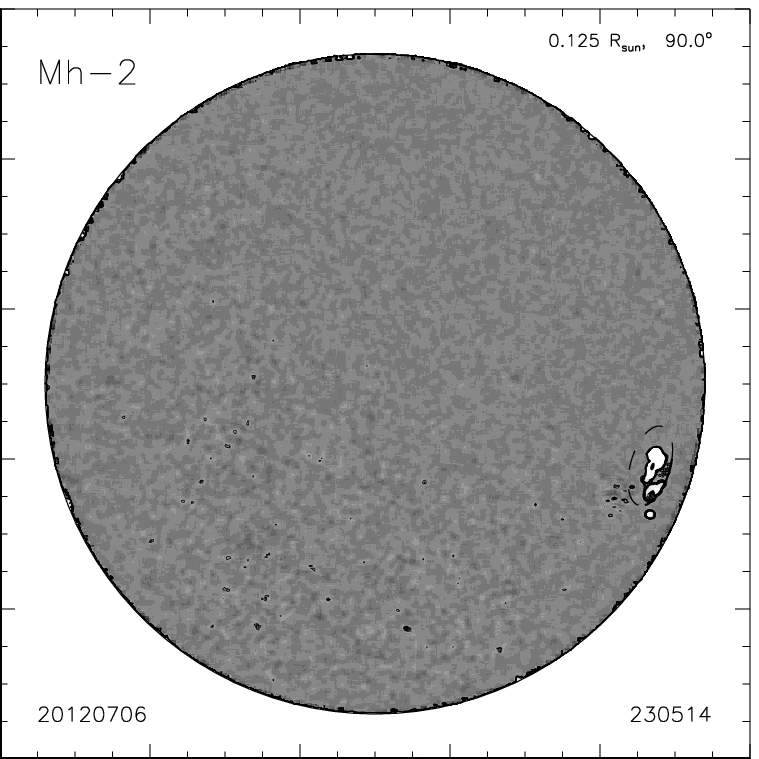}\hfill
\includegraphics[width=0.12\textwidth]{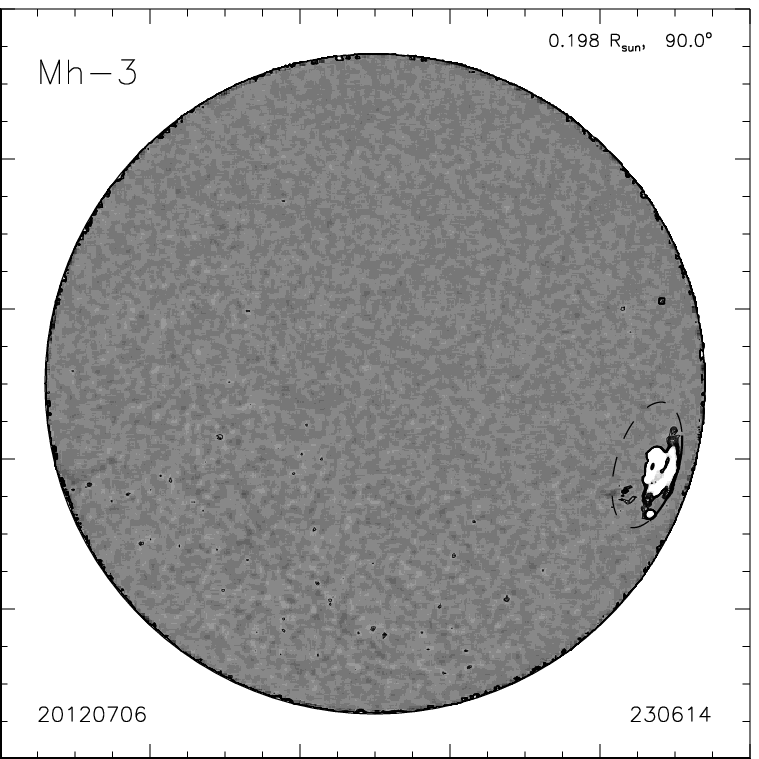}\hfill
\includegraphics[width=0.12\textwidth]{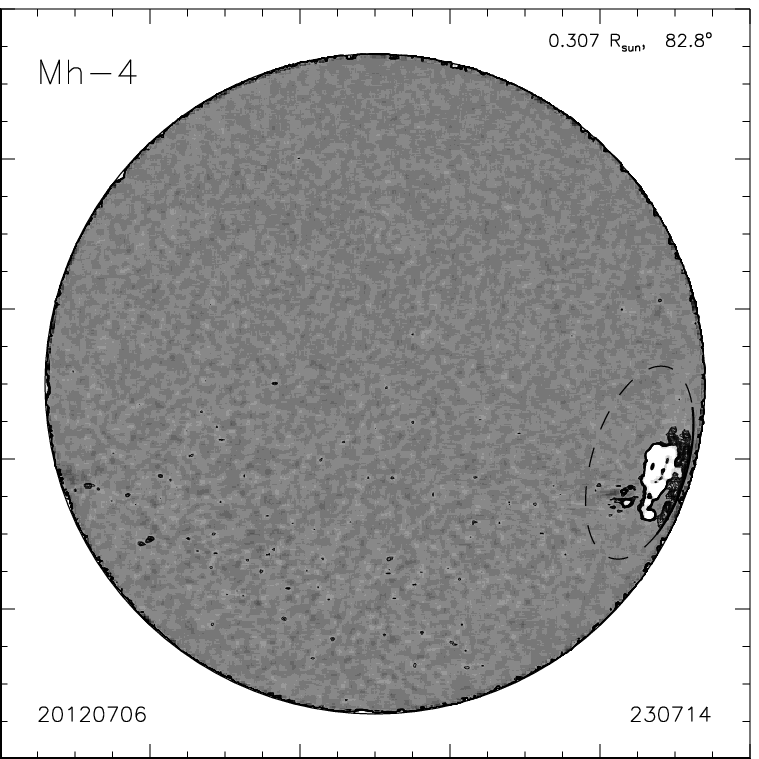}\\
\captionof{figure}{Similar with \autoref{halpha}, but for July 6 event observed in Mauna Loa Observatory.}
\label{halpha2}
\end{figure}

Among a number of x-ray flare detected by GOES since the begining of 24th Solar Cycle, we found two events that clearly show Moreton waves in series of GONG {\halp} images. The first event occured in June 3, 2012 between 17:52 - 17:57 UT after M3.3 flare in the active region NOAA 11496. During that time, four among six GONG observatories (Big Bear, Mauna Loa, Cerro Tololo, Teide) were able to record the central line {\halp} images with time resolution of $\sim20$ seconds. A total of 36 images are used for the kinematic analysis (see \autoref{halpha}). For the second event of July 6, there was an X1.1 flare in the active region NOAA 11515 occured at 23.01 UT, followed by near-limb Moreton wave observed from two GONG observatories (Big Bear, Mauna Loa). The minimum time resolution of the compiled images is $\sim20$ seconds (see \autoref{halpha2}).

For geometrical and kinematical analysis, different images are created by substracting the original images with the first image of interest, e.g. the image of 17:52 UT for June 3 event or 23:03 UT for July 6 event. This process is conducted independently for each observatory to accommodate different observational condition. After that, radial distance and angular extent of the wavefronts are measured manually by fitting circular arc centered on the wave origin (close to flare location).

\section{Result and Discussion}
\begin{figure}[ht]
\centering
\includegraphics[width=0.5\textwidth]{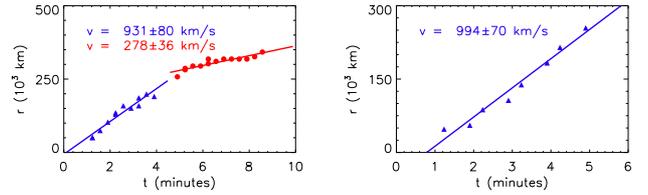}\\
\captionof{figure}{The kinematics of the leading edges of the Moreton waves observed in June 3 (left) and July 6 (right). Red symbols mark the kinematics of the brightening features in the north-eastern side of AR 1496.}
\label{speed}
\end{figure}

\begin{figure}[ht]
\centering
\includegraphics[width=0.45\textwidth]{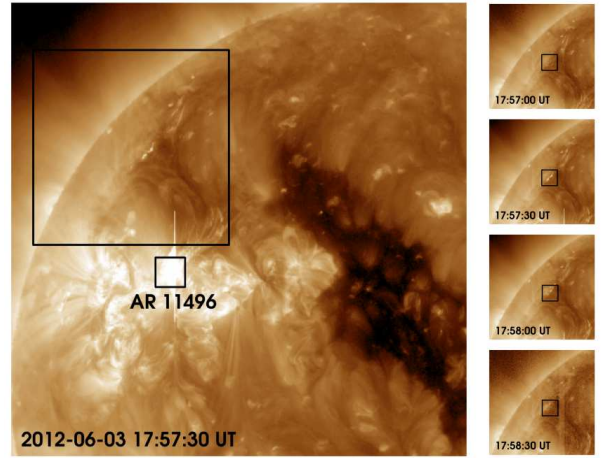}\\
\captionof{figure}{Left panels: AIA/SDO image of a part of the sun in June 3, 2012 17:57:30 UT. Right panels: series of the cropped images focus on the brightening feature which are co-spatially observed in {\halp} images.}
\label{eits}
\end{figure}

The Moreton wave of June 3, 2012 has a diffuse wavefronts that propagate to the north-west direction with a speed of $v=931\pm80$ km/s (\autoref{speed}). The broadness nature of this wave can be interpreted as the vertical extension of the wave over the chromosphere. Besides, we found a sudden brightening in the north-east of AR 1496 just after the main wavefronts fade away. This feature seems to move with significantly lower speed of $v=278\pm36$ km/s, a typical speed of EIT waves occur in the corona. The extreme ultraviolet images from Atmospheric Imaging Assembly/Solar Dynamics Observatory (AIA/SDO) also show co-spatial brightening (\autoref{eits}). The plausible interpretation of this brightening is a compression due to the shock wave from the flare. Because of the inhomogeneity of the plasma density in chromosphere and lower corona region, late brightening is observed instead of propagating wave from the flare location.

The Moreton wave of July 6, 2012 originates from AR 11515 which is located in the west limb of the sun at that time. The observed wave has angular span of about $90^{\circ}$. Propagation of the wave is $v=994\pm70$ km/s which is comparable to the derived particle velocity of type II burst observed from Culgoora Observatory. The location of the event makes it difficult to measure the geometry and kinematics precisely. In this case, wave deceleration is not clearly observed.

\acknowledgments
This work is supported by the Space Science Center, National Institute of Aeronautics and Space (LAPAN). RP gratefully acknowledge the travel grant from Faculty of Mathematics and Natural Sciences ITB and from IAU.


\begin{thebibliography}{}
\bibitem[\protect\astroncite{Balasubramaniam and {et al.}}{2010}]{bala10}
Balasubramaniam, K.~L. and {et al.}: 2010,
\newblock {\em ApJ} {\bf 723}, 587

\bibitem[\protect\astroncite{Chen and {et al.}}{2002}]{chen02}
Chen, P. F., Wu, S. T., Shibata, K., and Fang, C.: 2002,
\newblock {\em ApJ} {\bf 572L}, 99

\bibitem[\protect\astroncite{Hill and {et al.}}{1994}]{hill94}
Hill, F. and {et al.}: 1994,
\newblock {\em So. Ph.} {\bf 152}, 351

\bibitem[\protect\astroncite{Moreton}{1960}]{moreton60}
Moreton, G.~E.: 1960,
\newblock {\em ApJ} {\bf 65}, 494

\bibitem[\protect\astroncite{Temmer et~al.}{2009}]{temmer09}
Temmer, M., Vr\v{s}nak, B., \v{Z}ic, T., and Veronig, M.: 2009,
\newblock {\em ApJ} {\bf 702}, 1343

\bibitem[\protect\astroncite{Thompson and {et al.}}{1999}]{thompson99}
Thompson, B.~J. and {et al.}: 1999,
\newblock {\em ApJ} {\bf 517}, L151

\bibitem[\protect\astroncite{Uchida et~al.}{1973}]{uchida73}
Uchida, Y., Altschuler, M.~D., and Newkirk~Jr., G.: 1973,
\newblock {\em So. Ph.} {\bf 28}, 495

\bibitem[\protect\astroncite{Warmuth et~al.}{2004}]{warmuth04}
Warmuth, A., Vr\v{s}nak, B., Magdaleni\'c, J., Hanslmeier, A., and Otruba, W.:
  2004,
\newblock {\em A\&A} {\bf 418}, 1101
\end{thebibliography}
\end{document}